\documentclass[%
 reprint, superscriptaddress, 
 amsmath,amssymb, aps, prb, 
 ]{revtex4-2}
\usepackage{graphicx}
\usepackage{multirow}
\usepackage{comment}
\usepackage[usenames]{color}
\usepackage{graphicx}
\usepackage{dcolumn}
\usepackage{bm}
\usepackage[colorlinks=true, allcolors=blue]{hyperref}
\usepackage{amsmath}
\usepackage[normalem]{ulem}  

\newcommand{\lover}{l_{\tt o}}
\newcommand{\uover}{u_{\tt o}}
\newcommand{\ud}{{\tilde u}}
\newcommand{\hd}{h_{\tt d}}

\begin{document}


\title{Depinning free of the elastic approximation}

\author{A. B. Kolton}
\affiliation{Centro At\'omico Bariloche, CNEA, CONICET, Bariloche, Argentina}
\affiliation{Instituto Balseiro, Universidad Nacional de Cuyo, Bariloche, Argentina}

\author{E. E. Ferrero}
\affiliation{Instituto de Nanociencia y Nanotecnolog\'{\i}a, CNEA--CONICET, 
Centro At\'omico Bariloche, R8402AGP S. C. de Bariloche, R\'{\i}o Negro, Argentina.}
\affiliation{Departament de Física de la Matèria Condensada \& UBICS, Universitat de Barcelona, Martí i Franquès 1, 08028 Barcelona, Spain.} 

\author{A. Rosso}
\affiliation{Universit\'e Paris-Saclay, LPTMS, CNRS, 91405 Orsay, France}


\begin{abstract}
    We model the isotropic depinning transition of a domain-wall using a 
    two dimensional Ginzburg-Landau scalar field instead 
    of a directed elastic string in a random media.
    An exact algorithm accurately targets both the critical depinning field 
    and the critical configuration for each sample. 
    For random bond disorder of weak strength $\Delta$, the critical field 
    scales as $\Delta^{4/3}$ in agreement with the predictions for the 
    quenched Edwards-Wilkinson elastic model. 
    However, critical configurations display overhangs beyond a characteristic 
    length $l_{\tt 0} \sim \Delta^{-\alpha}$, with $\alpha\approx 2.2$, 
    indicating a finite-size crossover. 
    At the large scales, overhangs recover the orientational symmetry which 
    is broken by directed elastic interfaces. 
    We obtain quenched Edwards-Wilkinson exponents below $l_{\tt 0}$ and invasion 
    percolation depinning exponents above $l_{\tt 0}$. 
    A full picture of domain wall isotropic depinning in two dimensions is hence proposed.
\end{abstract}

\maketitle

In recent decades, significant progress has been made in understanding a paradigmatic 
example of out-of-equilibrium critical phenomena: the depinning transition of elastic 
interfaces in random media~\cite{Kardar1998,Nattermann2000,Fisher1998,Wiese2022}.
Depinning is relevant in various extended physical systems, such 
as ferromagnetic~\cite{Ferre2013} and ferroelectric~\cite{Kleemann2007,Paruch2013} 
driven domain walls (DWs), 
tensioned cracks in hard and soft matter~\cite{Bonamy2008,Ponson2009,LePriol2020},
the displacement of contact lines of liquid menisci~\cite{Joanny1984,Moulinet2004,LeDoussal2009}, or even stressed tectonic plates and earthquakes~\cite{Jagla2010,Jagla2014}. 
The common basic phenomenology of the depinning transition is well-captured 
by the simple model of a driven overdamped elastic interface 
coupled to a quenched disordered energy landscape that tends to trap 
the interface in configurations in which the potential energy is 
locally minimized.
Under the application of a uniform external driving force $f$ the
energy potential tilts.
The interface might move slightly, but if the amplitude of the driving 
force is below a well defined threshold $f_c$, it eventually pins and 
remains immobile.
Instead, it sets into a steady-state motion, with an average velocity $v>0$, 
if the driving force is above $f_c$. 
Exactly at $f_c$, the interface accommodates in a critical depinning 
configuration, exhibiting interesting universal geometrical properties.

Most progress in the field has been made by approximating the interface 
as {\it univalued} and using smooth scalar displacement fields for its dynamics, 
modeling it with overdamped equations like the driven quenched-Edwards-Wilkinson 
(qEW) model.
Through powerful analytical and numerical techniques, researchers have found 
that the depinning transition at $f_c$ is continuous, non-hysteretic, and occurs at 
a well-defined characteristic threshold force $f_c$~\cite{Kolton2013}.
At $f_c$ the interface is marginally blocked, and the instability is described 
by a localized soft spot or eigenvector~\cite{Cao2018}. 
Just above the threshold, the mean velocity $v$ follows the depinning 
law $v\sim (f-f_c)^{\beta}$, with $\beta$ being a non-trivial 
critical exponent~\cite{FerreroCRP2013}. 
A divergent correlation length $l \sim (f-f_c)^{-\nu}$ and a divergent 
correlation time $\tau \sim l^{z}$ characterize the jerky motion as 
$f_c$ is approached from above.
Below the length-scale $l$, the rough geometry of the interface becomes 
self-affine (SA), with the displacement field growing as $u \sim x^{\zeta}$ 
for length-scales $x$ below $l$. 
Hence $v\sim l^{\zeta-z}$ and $\beta=\nu(z-\zeta)$. 
Depinning critical exponents have been studied 
both analytically~\cite{LeDoussal2002,Fedorenko2003} and 
numerically~\cite{Leschhorn1993,Leschhorn1997,Roters1999,Rosso2003,Rosso2007,Ferrero2013,Ferrero2013b}.
Different universality classes are determined by the 
dimension of the interface $d$, the 
range~\cite{Ramanathan1998,Zapperi1998,Rosso2002,Duemmer2007,Laurson2013} 
or nature~\cite{Boltz2014} of the elastic interactions, 
the anisotropic~\cite{Tang1995} or isotropic correlations of the 
pinning forces~\cite{Fedorenko2006,Bustingorry2010}, 
and by the presence of additional 
non-linear terms~\cite{Amaral1994,Tang1995,Rosso2001b,Goodman2004,LeDoussal2003,Chen2015,Mukerjee2023,Mukerjee2023b}. 
If the so-called statistical tilt symmetry holds, 
only two exponents are needed to fully characterize 
the depinning universality class. 
At large velocities, the effect of disorder mimic thermal fluctuations, 
and $v \sim f$. 
For $f<f_c$, motion is only possible through thermal activation at 
a finite temperature $T$.
Particularly, for $f \ll f_c$ and relatively small temperatures, 
the universal creep-law~\cite{Ioffe1987,Nattermann1990,Chauve2000} 
$\ln v(f) \sim -f^{-\mu}/T$ holds,
with $\mu$ being a critical exponent related to dimension and the roughness 
exponent of the SA interface at thermal equilibrium ($f=0$).
Remarkably, in this ultra-slow creep regime, a depinning 
criticality emerges at large scales, as observed both 
in the steady-state geometry and in the spatio-temporal  fluctuations~\cite{Kolton2006,Kolton2009,Ferrero2017,Ferrero2021,Grassi2018}. 
Furthermore, several of these qEW predictions are quantitatively confirmed by
experiments conducted on ultra-thin ferromagnetic films with perpendicular 
anisotropy~\cite{Lemerle1998,Ferre2013,Jeudy2018,Jeudy2016,Grassi2018,Albornoz2021}.

However, the success of the elastic theory conflicts with the observation that,
in the same experimental systems, domain wall configurations often exhibit
{\it overhangs and pinch-off loops}, which, in fact, challenge the assumptions 
of the elastic theory.
The natural question that arises is: 
{\it Why, then, does the
theoretically minimalist approach of a purely elastic interface work
so well?
}


It is a common experimental practice to focus the analysis in a region displaying 
a ``well behaved'' univalued DW segment far from the nucleation centers and
``rare defects'' to test the theory. 
However, it is not clear whether overhangs are generated solely by rare 
strong pinning centers or extrinsic defects, as suggested by 
Ferre et al.~\cite{Ferre2013}, or whether they are generated 
by intrinsic defects that act cooperatively.
This type of intrinsic disorder is characterized by statistically uniform weak 
disorder with short-range correlations, as assumed in the elastic theory.
In other words, overhangs, fingers, and bubbles might be part of the solution 
of the actual critical interface in weak disorder and not an issue to be avoided.
Understanding why and to what extent the elastic theory applies is an important open question to gain a full understanding of the domain wall dynamics.

In this work, using a disordered scalar Ginzburg-Landau (GL)
model~\cite{Caballero2018,Caballero2020,Guruciaga2021,Caballero2021}, 
which does not break isotropy, we free ourselves from the elastic approximation, 
allowing for plasticity and realistic deformations of the interface.
With the help of an accurate algorithm, the critical fields and interface 
configurations can be solved at depinning for different realizations of the disorder.
Our analysis ultimately provides a comprehensive picture of DW depinning 
in two-dimensional isotropic media with short-range correlated disorder.
Critical configurations always display overhangs beyond a characteristic 
length $l_{\tt 0}$ that depends on the disorder strength $\Delta$, 
indicating a finite-size crossover. 
Below $l_{\tt 0}$, we obtain quenched Edwards-Wilkinson exponents,
and above $l_{\tt 0}$, we observe invasion percolation depinning exponents.

It's worth mentioning that other approaches to the isotropic depinning 
transition have been pursued.
Numerical simulations of the random-field Ising model (RFIM) suggest that 
in two dimensions, the DW critical configurations at the depinning threshold 
should be either faceted or self-similar (SS), instead of the self-affine (SA) 
geometry predicted by the elastic theory~\cite{Ji1991,Koiller1992}.
However, these simulations are unable to address the isotropic depinning 
transition in the relevant weak disorder case due to the coupling of their 
thin domain walls to the underlying periodic lattice~\cite{Koiller1992}, 
which breaks isotropy.

The paper is organized as follows:
Sec.\ref{sec:model} introduces the model.
Sec. \ref{sec:algo} presents the algorithm that we developed to target 
the critical configurations at depinning. 
In Sec.\ref{sec:results}, we describe and discuss our numerical results: 
first, the critical field (Sec. \ref{sec:hd}) and then the geometry of the 
critical configurations (Sec.\ref{sec:configs}).
In Sec.~\ref{sec:conclusions}, we summarize our conclusions. 
The appendix contains further details of the numerical computations.


\section{Model}
\label{sec:model}
We consider the following time-dependent Ginzburg-Landau equation for a scalar order parameter field $\phi(x,y)$, describing approximately the zero-temperature evolution of the out of plane magnetization in a ferromagnetic film with perpendicular anisotropy~\cite{Caballero2018,Caballero2020,Guruciaga2021,Caballero2021},
\begin{equation}
\gamma \partial_t \phi = c\nabla^2 \phi + 
\epsilon_0[(1+r(x,y)) \phi - \phi^3] + {h}.
\label{eq:phi4}
\end{equation}
Here, the elastic constant $c$, $\epsilon_0$ and the friction $\gamma$ 
are positive constants, $h$ is the constant and uniform magnetic field,
and $r(x,y)$ is a quenched random bond disorder specified by 
\begin{eqnarray}
\overline{r(x,y)}&=&0, \nonumber \\
\overline{r(x',y')r(x,y)}&=& \frac{2}{3} \Delta^2 \delta(x-x')\delta(y-y'),
\label{eq:disorder}
\end{eqnarray}
We denoted $\overline{(...)}$ the average over disorder realizations and  
$\Delta$ the disorder strength. 
We are interested in the limit of small disorder where the ground state 
at zero field is ferromagnetic 
(i.e. either $\phi\approx 1$ or $\phi \approx -1$). 

We prepare the system with a small positive domain, namely $\phi \approx 1$ 
for $y<y_0$ and $\phi=-1$ for $y>y_0$.  
After an initial transient at $h=0$ we observe a well defined domain wall 
of width $\delta \approx \sqrt{c/\epsilon_0}$ and surface tension 
(DW energy per unit length) $\sigma \approx \sqrt{c \epsilon_0}$. 
The location of the wall is defined as the level zero set of the magnetization, 
i.e. the set of points $(x^*,y^*)$ where $\phi(x^*,y^*)=0$.
Upon the application of a magnetic field $h$ (that favors the positive
$\phi$ phase) the domain wall would acquire a steady velocity $v(h)=m h$, 
with $m \approx \delta/\gamma$ the mobility of the wall;
but in the presence of disorder, the velocity-field characteristics 
becomes non-linear and displays a depinning transition: 
above a threshold field $h_{\tt d}$ the DW slides at a finite velocity, 
while below it the wall gets trapped into one of many metastable states.
This has been clearly observed and quantified even in these scalar GL models 
where interfaces are not forced to be univalued~\cite{Caballero2018}.
Our goal is to determine $h_{\tt d}$, and the last metastable state or 
critical configuration $\phi_d(x,y)$, as $h \to h_{\tt d}$ from below.

\section{Algorithm}
\label{sec:algo}

Eq.(\ref{eq:phi4}) is discretized with  a regular $L \times L$ grid with coordinates $(i,j)$. 
\begin{eqnarray}
\label{eq:discrete}
\partial_t \phi_{i,j} 
&=& c[\phi_{i+1,j}+\phi_{i+1,j}+\phi_{i,j+1}+\phi_{i,j-1}-4\phi_{i,j}] \nonumber \\
&+& \epsilon_0[(1+r_{i,j}) \phi_{i,j} - \phi_{i,j}^3] + {h}=0,
\end{eqnarray}
Here, $r_{i,j}$ are  uncorrelated random numbers sampled from a uniform distribution in 
$[-\Delta,\Delta]$. 
Setting $0<\Delta<1$,  the term proportional to $\epsilon_0$ 
in Eq.(\ref{eq:discrete}) admits always three zeros for $|\phi_{i,j}|<1$.
We apply periodic boundary conditions in the $x$ direction $\phi_{i+L,j}=\phi_{i,j}$, 
but anti-periodic boundary conditions in the $y$ direction, $\phi_{i,L}=-\phi_{i,0}$ 
to ensure that the evolution affect a single domain wall.

{Starting from an arbitrary initial configuration 
(typically with a flat interface),
we evolve the scalar field $\phi_{i,j}$ according to the so-called
``variant Monte Carlo'' method~\cite{Rosso2001b,Rosso2003,Rosso2005}.}
An elementary move updates $\phi_{i,j}$ by replacing it to one of the roots  
of Eq.(\ref{eq:discrete}) with $\partial_t \phi_{i,j} =0$. 
Using the François Vi\`ete formula for the cubic equation we find at most 
three real roots. 
The root that we select is the closer to the initial position $\phi_{i,j}$. 
The elementary steps are repeated consistently over the sites $(i,j)$,
until the (positive) mean velocity of 
the order parameter is smaller than a small cutoff $\epsilon=10^{-4}$, 
signaling the proximity to a metastable state. 

\begin{figure}[t!]
\includegraphics[width=\columnwidth]{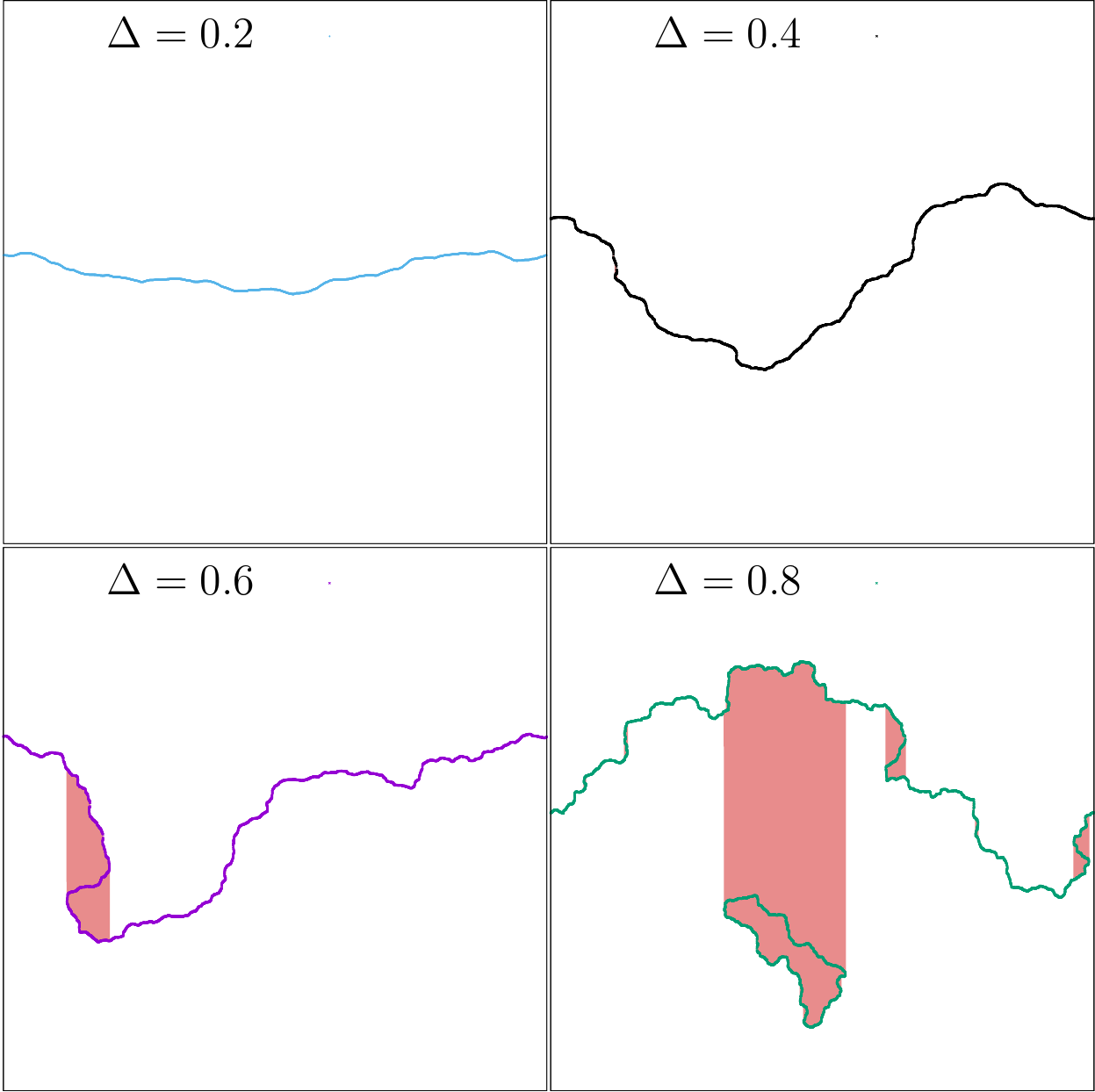}
\caption{
Typical critical configurations of the domain wall for different values of the 
disorder strength $\Delta$. 
System size is ${L=4096}$ (the aspect ratio is one). 
Red shaded regions highlight overhangs and/or pinch-off loops 
}
\label{fig:samplesnew}
\end{figure}

This strategy relies on the Middleton theorems
The ``no-passing theorem'' assures that the algorithm connects an arbitrary initial state with the critical configuration, whereas the ``forward-moving theorem'' justifies the approach of making only forward directed elementary moves~\cite{Rosso2005}. 
This allows us to use the same technique developed in 
Refs.~\cite{Rosso2001b,Rosso2003,Rosso2005} for the elastic interface. 
The novelty of our implementation is that instead of applying it to the 1D DW 
displacement field, we apply it directly to the 2D order-parameter $\phi_{i,j}$. 
The order parameter is always univalued and bounded and has a convex elastic energy, though it can nucleate the multivalued critical DW we are more generally 
interested in.
The algorithm proposed is, for a given accuracy, faster than the actual 
dynamics of Eq.(\ref{eq:phi4}); because elementary moves are not 
proportional to any numerical integration time-step, but instead 
controlled by a root finding 
method~\footnote{
The computational cost of the present algorithm is equivalent to the 
one applied to two-dimensional univalued interfaces (only the nonlinear 
force changes). 
We refer the reader to Ref.~\cite{Rosso2003} for a discussion about its
performance compared with the direct integration of the equation of motion 
of the 2d interface (i.e. the equivalent to Eq.(\ref{eq:phi4})).
}. 
We thus alleviate the critical-slowing down near all metastable states.

Furthermore, many of the computations involved in the implementation of the algorithm 
can be performed simultaneously, thus offering a valuable opportunity for an 
optimization via parallelization. 
These computations, such as elementary moves using the checkerboard decomposition, 
DW detection and other image processing like routines and reductions to obtain 
properties are accelerated using graphics processors. 
To apply the algorithm we discretize the two dimensional space and use finite 
differences to evaluate the derivatives. 
We consider square systems of size $L\times L$. 
Without loss of generality we choose $\epsilon_0=1$, $c=1$, $\gamma=1$, 
and for each disorder realization we drew $L\times L$ uncorrelated numbers 
$r(x,y)$ from a uniform distribution. 
The correlation length of the disorder $r_{\tt f}$ is thus of the order of the 
discretization itself and smaller than the DW width which then becomes 
the correlation length of the pinning force on DWs.

\begin{figure}[t!]
\includegraphics[width=\columnwidth]{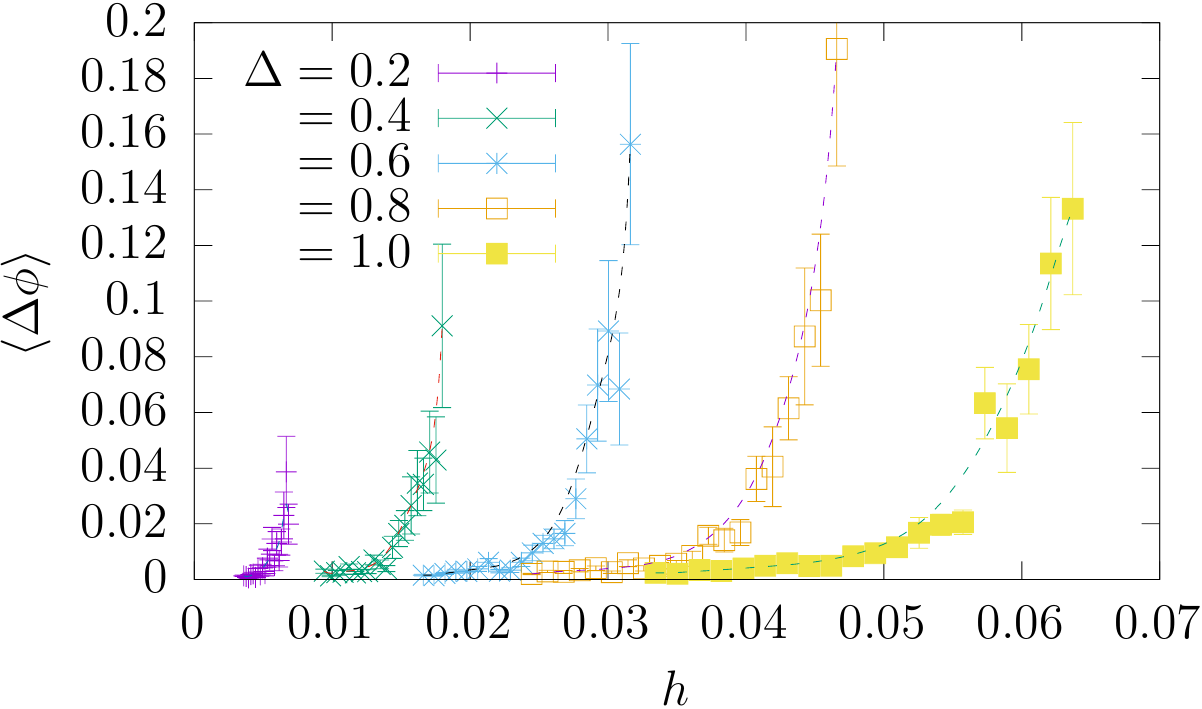}
\caption{
Magnetization jump per site $\langle \phi \rangle$ as a function of $h$, 
for different values of the disorder strength $\Delta$. 
Data correspond to averages over 50 samples, $L=256$.
Dashed-lines are guides to the eye. 
}
\label{fig:howtogethd}
\end{figure}

We build a sequence of metastable states as a function of $h$ until we localize 
$h_{\tt d}$, above which no metastable state can be found in the sample. 
Typical critical configurations for different values of $\Delta$ are shown in Fig.\ref{fig:samplesnew}. 
Overhangs can be clearly identified in the shaded-areas 
(defined as the regions where the interface adopts more than one value in the
advancement direction) and they are more frequent for large disorder. 
Interestingly, for the largest disorder we can also see isolated pinned pinch-off 
loops signaling domains that were not flipped.

The process of searching the root can be accelerated by a bisection method in the 
variable $h$. 
The critical field $h_{\tt d}$ can be thus obtained with the desired precision but 
with a price: the closer $h_{\tt d}$ the larger the average simulation time due to 
occurrence of large avalanches~\cite{Kolton2006b}. 
To illustrate this we computed the magnetization jumps 
\begin{equation}\label{eq:deltaphi}
    \langle \Delta \phi \rangle := 
\overline{
\frac{1}{L^2}\sum_{i,j} 
\left[
\tilde\phi_{i,j}(h+\delta h)-\tilde\phi_{i,j}(h)
\right].
}
\end{equation}
where $\tilde \phi_{i,j}(h)$
refers to the metastable configuration obtained at a given $h<h_d$.

In Fig.~\ref{fig:howtogethd} the divergences for different values of 
$\Delta \phi$ are signatures of the $h_{\tt d}$. 
The increase of $h_{\tt d}$ with increasing $\Delta$ can be appreciated. 
Notice that one could divide Eq.\ref{eq:deltaphi} by $\delta h$ 
and interpret it as a generalized susceptibility 
$\chi \equiv \frac{\Delta \phi}{\delta h}$.
Yet, its quantitative comparison with a true susceptibility
in experiments would be meaningless at this point.

\section{Results}
\label{sec:results}

Depending on the system size $L=128,256,\dots,8192$, we obtain from hundreds 
to thousands critical configurations of the DW, each one with their respective 
critical field $h_{\tt d}$ with an accuracy of $\Delta h_{\tt d} = 10^{-4}$. 
This sampling allow us to perform statistics and analyze several properties. 
In Sec.\ref{sec:hd} we discuss the average over disorder of the critical 
field $h_{\tt d}$ for different values of the disorder $\Delta$. 
In Sec.\ref{sec:configs} we discuss the geometry of the obtained 
critical configurations.

\subsection{Critical Field}
 \label{sec:hd}

\begin{figure}[t!]
\includegraphics[width=\columnwidth]{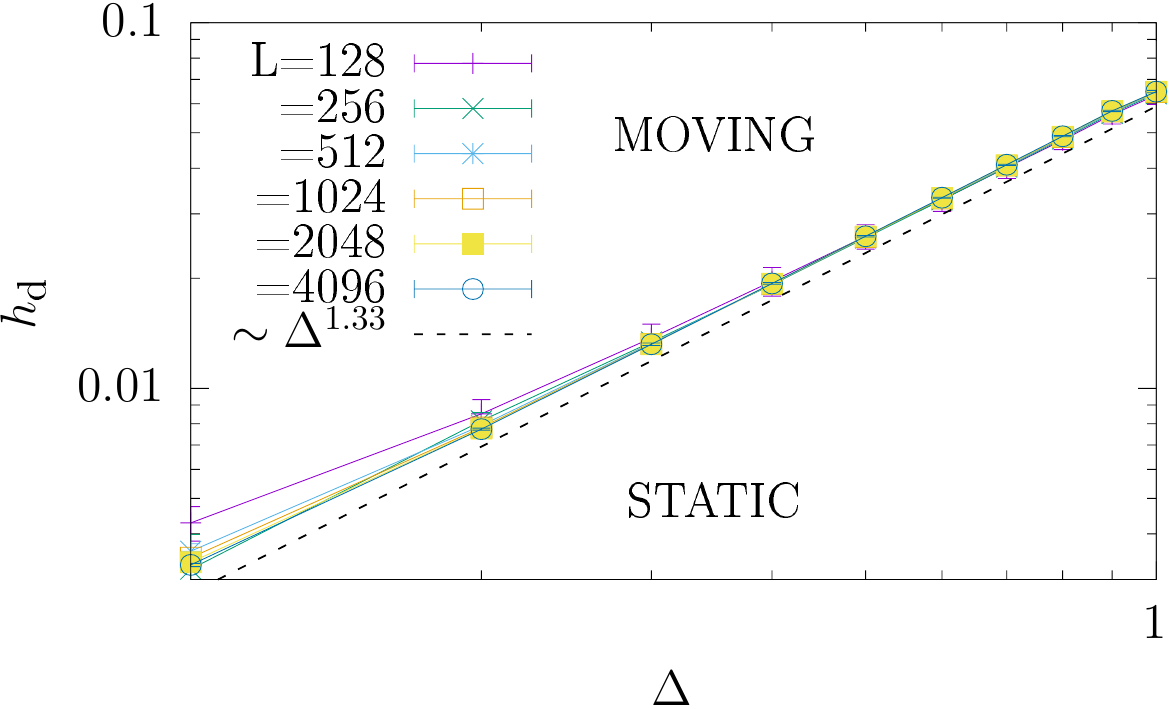}
\caption{Phase diagram of the domain wall dynamics.
A critical field $h_{\tt d}$ as a function of disorder strength $\Delta$ separates 
the `static' and `moving' phases.
It displays a $h_{\tt d} \sim \Delta^{1.33}$ in agreement with the predicted behavior
$h_{\tt d} \sim \Delta^{4/(4-d)} (\sigma\delta)^{-d/(4-d)}$ of weak collective pinning 
theory (dotted line). 
We display it for different system sizes $L$, finite-size effects are only observed 
for very small $\Delta$.
}
\label{fig:hdvsDelta}
\end{figure}

We compute the average critical field $h_{\tt d}$ as a function of $\Delta$ for different sizes $L$.
Fig.\ref{fig:hdvsDelta}  shows the power law dependence  $h_{\tt d} \sim \Delta^{4/3}$. 
This result  is consistent with Larkin's prediction for weak collective pinning where $h_{\tt d}\sim {\sigma \delta l_{\tt c}^{-2}}$, with the Larkin length $l_{\tt c} \approx \left({\sigma \delta/\Delta} \right)^{2/(4-d)}$ (note that the microscopic pinning correlation length is the DW width $\delta$). Finally we have \cite{Blatter1994}:

\begin{equation}
h_{\tt d} \sim \Delta^{4/(4-d)} (\sigma\delta)^{-d/(4-d)},  
\label{eq:Larkin}
\end{equation}
with  $d=1$ in our case. 
The deviations observed for small $\Delta$ and $L$ can also be explained from the Larkin theory, since for small $\Delta$, $l_{\tt c}$ becomes of the order of $L$. In this case,
$l_{\tt c}$ is replaced by $L$, and thus $h_{\tt d}\sim{\sigma \delta L^{-2}}>{\sigma \delta l_{\tt c}^{-2}}$. 
The fact that the weak collective pinning theory is applicable shows that, at least 
up to the scale $l_{\tt c}$, the DWs can successfully be described as a smooth elastic interface with univalued displacement field. 
It also shows that grid-pinning effects are negligible and that depinning is isotropic for the whole $\Delta$ range analyzed (see a detailed discussion in Appendix~\ref{sec:circularity}).

\begin{figure}[t!]
\includegraphics[width=\columnwidth]{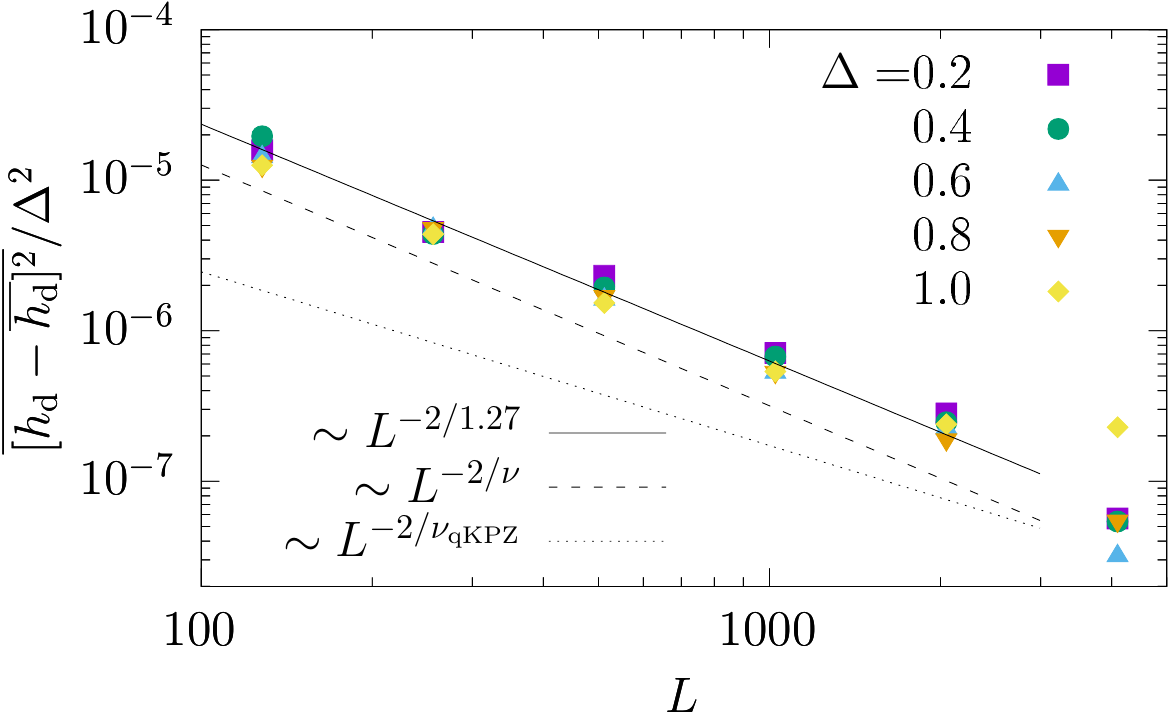}
\caption{
Size dependence of the sample to sample fluctuations of the critical field for 
different disorder strengths $\Delta$.
Data is fairly described $\overline{[h_{\tt d}-\overline{h_{\tt d}}]^2}\sim \Delta^2 L^{1/1.27}$ (solid line), as can be appreciated by fitting the rescaled data. 
For comparison we plot the scaling expected for the qEW model (dashed-line) and 
for the qKPZ model (dotted-line).
}
\label{fig:hdflucvsL}
\end{figure}

We have also analyzed the sample to sample fluctuations of the depinning field $h_{\tt d}$. 
These fluctuations are expected to scale with the size $L$ of the interface as
\begin{equation}
    \overline{[h_{\tt d}-\overline{h_{\tt d}}]^2}\sim L^{-2/\nu},
    \label{eq:sampletosamplehd}
\end{equation}
For the 1d-qEW model of a SA interface we expect $\nu\approx 4/3$. 
Quite surprisingly (or maybe not), the same exponent is expected for 
invasion percolation $\nu^{\tt IP}\approx 4/3$.
In Fig.\ref{fig:hdflucvsL} we plot the critical field fluctuations as a 
function of $L$ for different values of $\Delta$. 
The best fit (solid-line) yields 
$\overline{[h_{\tt d}-\overline{h_{\tt d}}]^2}\sim \Delta^2 L^{-2/(1.27\pm 0.1)}$, 
fairly close to $\nu\approx 4/3$ (dashed-line) expected for the 1d-qEW model~\cite{Ferrero2013}, the invasion percolation model~\cite{Maslov1995}
and also to the one observed in 2D-RFIM simulations with strong disorder~\cite{Ji1991}. 
In fact, we will argue that this is not just a coincidence.

\subsection{Critical Configurations}
 \label{sec:configs}

We now describe the geometrical properties of the configurations corresponding 
to the critical fields discussed in the previous section. 

\begin{figure}[t!]
\includegraphics[width=\columnwidth]{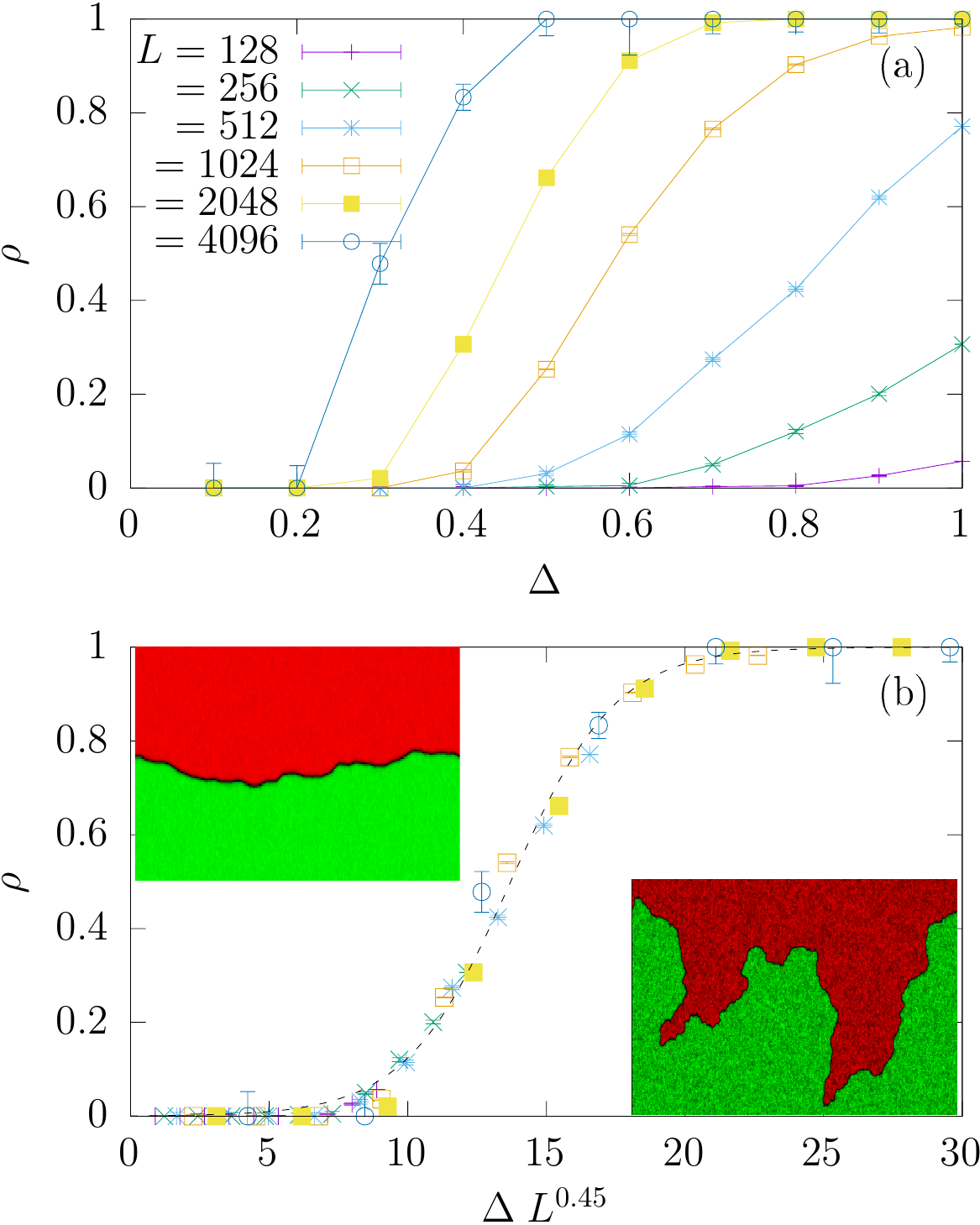}
\caption{
Fraction $\rho$ of multivalued domain walls at the depinning transition. 
{\bf (a)} Raw data as a function of disorder strength $\Delta$ and transverse size $L$. 
{\bf (b)} Scaled data, showing that $\rho(\Delta,L) \approx
\tilde{\rho}(L/\lover)$ with $\lover \sim \Delta^{-1/0.45}$. 
The inset shows typical domain walls for $L<\lover$ (left) and $L>\lover$ (right).
}
\label{fig:overhangs}
\end{figure}

\subsubsection{Overhangs}

For a fixed $L$, depending on $\Delta$, we find that there is a finite probability that the critical configuration presents overhangs (see insets of Fig.\ref{fig:overhangs}), and also pinch-off loops (see Fig.\ref{fig:samplesnew}). 
These are detected as multivalued DW displacements with respect to the reference flat initial configuration.  
We define the overhang probability $\rho$ by counting, for a fixed $L$ and $\Delta$,  the fraction of critical DW that presents overhangs for many randomness realizations. 
In Fig.\ref{fig:overhangs}(a) we see that always $\rho \to 1$ for large $\Delta$, and $\rho \to 0$ for small $\Delta$. 
Interestingly, the crossover depends on $L$, and the {\it empiric} scaling law 
$\rho(\Delta,L) \approx \tilde{\rho}(L/\lover)$, with 
\begin{equation}
\lover \sim \Delta^{-1/0.45}, 
\end{equation}
decently fits the data for all the $L$ and $\Delta$ values considered, as shown in Fig.\ref{fig:overhangs}(a). 
The success of this fit indicates that there exists a crossover length associated to the average overhang size, $\lover \sim \Delta^{-1/0.45}$. 
Since $\lover$ diverges when $\Delta\to 0$ and not in a finite value 
of $\Delta$, our results indicate that there is no transition to a phase 
of univalued interfaces but a disorder-driven crossover instead, and that 
DWs are always expected to present overhangs in the thermodynamic limit. 


A simple heuristic argument can be used to relate the super-roughness predicted 
for the qEW model and the occurrence of overhangs. 
If we assume overhangs appear at the scale $l$ where the extrinsic DW width or ``global roughness'' $w$ satisfy $dw/dl \sim 1$ with $w 
\approx \delta (l/l_{\tt c})^{\zeta}$ and $\zeta$ the roughness exponent at that scale, 
we get 
\begin{equation}
\lover \sim (l_{\tt c}^{\zeta}/\delta)^{1/(\zeta-1)}\sim \Delta^{-\frac{2\zeta}{(4-d)(\zeta-1)}}
\label{eq:lover}
\end{equation}
Using that in $d=1$  $l_{\tt c} \sim \Delta^{-2/3}$ we need $\zeta \approx 1.43$ 
to obtain the observed exponent of $-1/0.45 \sim -2.2$ in the scaling of $\lover$ 
with $\Delta$. 
The  value for this effective $\zeta$ remains to be explained, 
but lies in between the $d=1$ Larkin exponent $\zeta_{\tt L}=3/2$ and 
the  $d=1$ depinning exponent $\zeta \approx 5/4$. 
More importantly, we note that, to obtain a finite $\lover > l_{\tt c}$, the argument 
requires super-roughening ($\zeta>1$) above $l_{\tt c}$, and suggests that overhangs 
may not affect the SA geometry if $\zeta<1$ (and for moderate $\Delta$). 
This may explain why SA two-dimensional interfaces are observed in a three dimensional 
space at depinning~\cite{Clemmer2019}, using that $\zeta \approx 0.75$ for the 
2d-qEW~\cite{Rosso2003}. 
The argument then predicts one dimensional SA critical interfaces in the long-range 
elasticity isotropic depinning case for which $\zeta \approx 0.39$~\cite{Rosso2002,Duemmer2007}. 
In order to test the scaling argument it would be interesting to address
these predictions directly in future work. 

\subsubsection{Roughness exponent}

In the previous sections we have shown that DWs at the isotropic depinning 
transition have always overhangs in the thermodynamic limit. 
The existence of a large crossover length $\lover \sim \Delta^{-2.2} \gg l_{\tt c}$ 
at weak disorder suggests that the random-manifold regime of the elastic-theory may
nevertheless exist at intermediate scales, where the probability of having one or 
more overhangs in a configuration is low. 
To test such an hypothesis we next focus on the roughness of the critical configurations. 

Standard methods to estimate the roughness exponent $\zeta$ of an interface rely on the existence of a univalued displacement field $u(x)$. 
We have shown however the interface at a given $x$ can be multievaluated, $\{u_n(x)\}_{n=1}^{n=m(x)}$, with $m(x)=1,3,5,\dots$.
Hence we define a univalued displacement field 
\begin{equation}
\ud(x):= \frac{1}{m(x)}\sum_{n=1}^{m(x)} u_n(x).   
\label{eq:ud}
\end{equation}
This particular choice coincides with the usual univalued interface if $m(x)=1\;\forall x$,
but in the presence of overhangs it introduces artificial discontinuities in $\ud(x)$. 
Besides this warning, it is an operationally well defined regularization, not only for simulation but also for experiments.
A convenient  way to obtain the roughness exponent at different length scales  
is through the structure factor 
\begin{equation}
    S(q) :=\overline{\left|\sum_{x=1}^L  e^{iqx} \ud(x)\right|^2} \sim \frac{1}{q^{1+ 2 \zeta}}.
\end{equation}  

\begin{figure}[t!]
\includegraphics[width=\columnwidth]{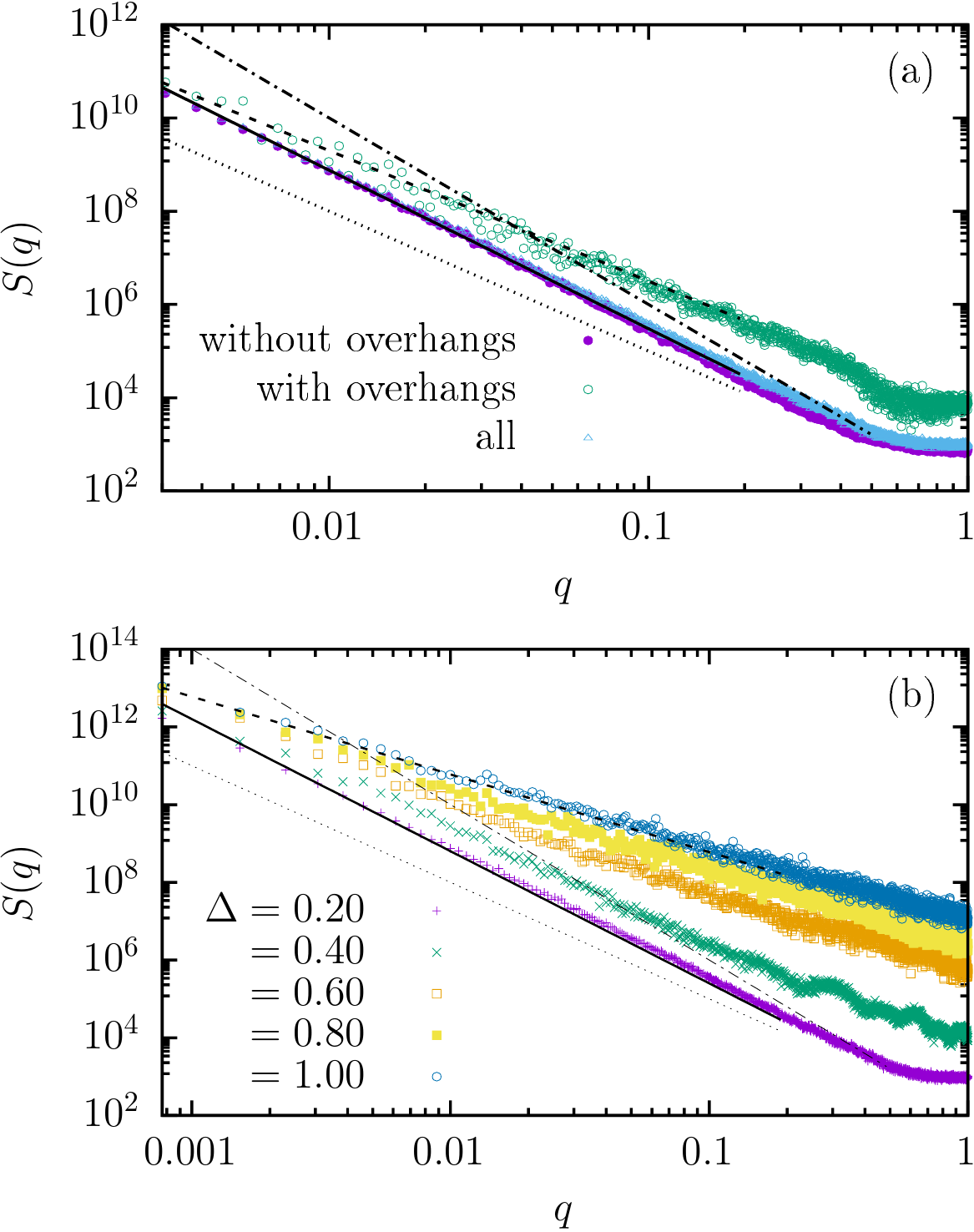}
\caption{Disorder averaged structure factor $S(q)$ of domain wall critical configurations.
Data obtained for a system size $L=8192$ (corresponding to a grid of 
${8192^2 \simeq 6.7\times 10^7}$ sites).  
{\bf (a)} 
$S(q)$ averaged over configurations with overhangs, without overhangs, 
and averaged over all configurations (180 samples), for a fixed disorder 
strength $\Delta=0.2$.
Lines are guides to the eye and display pure power-laws for $\zeta=1.2$ (solid-line) $\zeta=1$ (dotted-line), and $\zeta=0.9$ (dashed-line).
{\bf (b)} $S(q)$ averaged over all sampled critical configurations for increasing values of 
disorder strength $\Delta$ (180 samples for $\Delta=0.2$, 50 samples for $\Delta\!>\!0.2$). 
Guides-to-the-eye lines display pure power-laws for $\zeta\!=\!1.2$ (solid-line) and $\zeta\!=\!0.5$ (dashed-line).
}
\label{fig:weak8192}
\end{figure}

In Fig.\ref{fig:weak8192}(a) we show $S(q)$ for $\Delta=0.2$ and $L=8192$. 
For these parameters $\rho\approx 0.2$, so -on average- one over five critical
configurations is expected to have one or more overhangs.
If we separate the contributions from interfaces with and without overhangs we 
can observe marked differences. 
For low $q$,  in the case without overhangs, we can very accurately fit an exponent 
$\zeta \approx 1.2$  (solid-line) which is compatible with the reported values 
$\zeta \approx 1.25$ for the 1d-qEW at depinning~\cite{Ferre2013}. 
In the case with overhangs we can reasonably fit (dashed-line) an effective exponent
$\zeta_{\tt eff}=0.9$ over the low-$q$ region, even if overhangs have an important 
effect for the whole range of $q$. 
For the chosen value $\Delta=0.2$, the average over all configurations 
(with and without overhangs) is dominated by the ones without overhangs. 
Nevertheless, as we increase $\Delta$ (or increase $L$) the configurations 
with overhangs dominate the average. 
In Fig.\ref{fig:weak8192}(b) we show the results for increasing values of $\Delta$, 
the value of   $\zeta_{\tt eff}$ decreases with increasing $\Delta$ and tends to 
an exponent $\zeta_{\tt eff}\approx 0.5$ 
\footnote{It is worth noting that the decrease of $\zeta_{\tt eff}$ with increasing 
$\Delta$ is similar to the behavior reported in~\cite{Qin2012} for the spectral 
roughness exponent in the 2D-RFIM using the Monte Carlo method.}.

The results of Fig.\ref{fig:weak8192} support the applicability of the 1d-qEW model to describe DW at depinning to scales below $\lover$, where overhangs are rare. 
On the other hand,  overhangs  produce lower effective roughness exponents. 
These must be interpreted carefully because  they depend on our particular 
way to define $\ud$ in Eq.(\ref{eq:ud}). 
This result thus stresses the importance of a conscious interpretation of 
experimental data when $L > \lover$. 
More importantly, it stresses the importance of estimating quantitatively 
the crossover length $\lover$, both in experiments and simulations at the 
isotropic depinning transition. 

\subsubsection{Overhangs size scaling}

\begin{figure}[b!]
\includegraphics[width=\columnwidth]{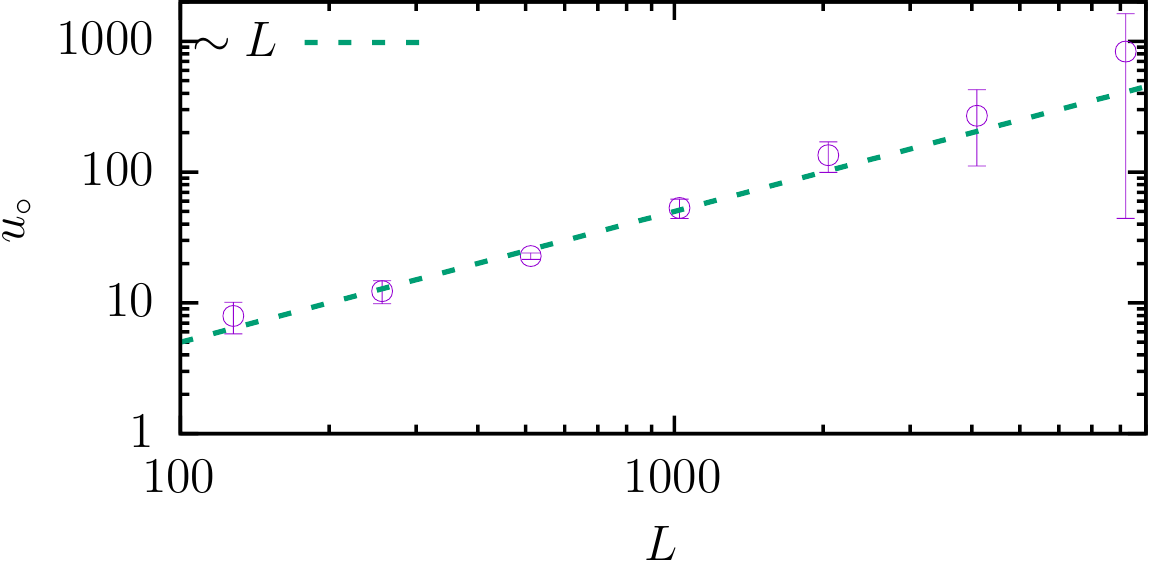}
\caption{Typical linear size of overhangs $\uover$ (Eq.(\ref{eq:overhangsize})) in the displacement direction as a function of the system size $L$ for $\Delta=1$. The dashed-line is a guide to the eye with slope $\propto L$. 
}
\label{fig:overhangsize}
\end{figure}

In Fig.\ref{fig:weak8192}(b) we can see that for strong disorder 
$\zeta_{\tt eff}\approx 0.5$, compatible with simple thermal roughening. 
One can interrogate whether
this result is an artifact introduced by the regularization of Eq.(\ref{eq:ud}) 
or rather an indication that a SA interface is emerging in spite of the presence 
of overhangs.
In order to answer this question, with our model we analyze how the typical size 
of overhangs scales with the  system linear size $L$.
To do this we first define
\begin{equation}
    \uover^{2} := \overline{
    \frac{1}{L}\sum_{x=1}^L 
    \left[ 
    \sum_{n=1}^{m(x)} \frac{u_n(x)^2}{m(x)} -
    \left(\sum_{n=1}^{m(x)} \frac{u_n(x)}{m(x)}\right)^2\right]}.
\label{eq:overhangsize}
\end{equation}
The quantity between brackets $[\dots]$ is identically zero if there is no overhang 
at $x$ (since in that case $m(x)=1$), and it is of the order of the overhang size 
squared otherwise (i.e. the variance of the multiple values of $u(x)$). 
Therefore, $\uover$ gives the typical size of the overhangs in a given configuration 
(note that for a univalued function $\uover=0$ regardless of its roughness).

In Fig.~\ref{fig:overhangsize} we show that for $\Delta=1$, 
$\uover$ scales approximately linearly with $L$. 
This demonstrates that overhangs are relevant in the thermodynamic limit 
and can not be eliminated by coarse-graining.
The snapshot shown in the right side inset of Fig.\ref{fig:overhangs} illustrates 
this result for a particular critical configuration.
Therefore, the roughness exponent $\zeta_{\tt eff}$ cannot be related to any 
emerging SA scale invariant.
On the other hand, this suggests that overhangs allow to recover the rotational symmetry, 
as they promote wandering of the local orientation of the DW.
So that the final picture for the roughness of the DW is:
SA qEW $\to$ finite size crossover $\to$ SS (invasion percolation depinning), 
as larger length scales are tested.

\section{Conclusions}
\label{sec:conclusions}

In summary, using an accurate algorithm we showed that critical configurations at the
isotropic depinning transition of one dimensional DWs always present thermodynamically 
large overhangs in the thermodynamic limit. 
Rotational invariance is thus not broken at the isotropic depinning transition of $d=1$
dimensional DWs in $d=2$ random media. 
We find nevertheless a crossover length below which the predictions of the elastic theory 
of univalued interfaces predictions are well satisfied, including the qEW super-roughness.
This extends the current theoretical understanding of depinning and reconciles it with 
many experimental observations where compatibility with the predictions for the 1d-qEW depinning universality class is found.
We thus propose a full picture for the DW depinning in isotropic two dimensional media 
which we hope will motivate further experimental and theoretical research on the 
depinning transition.

Our results could be relevant for experiments. 
While so far they can only compare at a qualitative level, the long standing 
statistical physics questions addressed in our work are important even for 
the experimental protocol conception and data analysis. 
For instance, work on magnetic DWs in the lab sometimes avoid including in the
analysis non-univalued pieces of interface and/or disregard pinch-offs as 
rare impurities related occurrences. 
We could now argue that those are not things to avoid, that they make part 
of the isotropic depinning theory and play a role in preserving the DW characteristics.
We then hope that our work could motivate a new look to experimental data 
(even the already existent one), perhaps not only in magnetic domain walls 
but also on different experimental realizations of 1D DWs in 2D
random media.

\begin{acknowledgements}
We thank F. Paris for early discussions.  
We acknowledge support from the CNRS IRP Project ``Statistical Physics of Materials'', 
PICT2019-1991 (MinCyT), and UNCuyo C014-T1 
and PIP 2021-2023 CONICET Project Nº 0757. 
ABK acknowledges hospitality at the LPTMS group where this project
started a few years ago.
EEF acknowledges support from the Maria Zambrano program of the 
Spanish Ministry of Universities through the University of Barcelona,
and MCIN/AEI support through PID2019-106290GB-C22.
We have used computational resources from CCAD-UNC and GF-CNEA which 
are part of SNCAD-MinCyT, Argentina.
\end{acknowledgements}

\appendix

\section{Control of lattice effects}
\label{sec:circularity}

In order to solve Eq.(\ref{eq:phi4}) numerically, 
a finite difference scheme based on an anisotropic regular 
lattice is typically used.
If one wants to simulate isotropic depinning, it is important to 
assure that the effect of the discretization is negligible. 
This effect can be greatly reduced and put under control by 
choosing the DW width $\delta$ large enough. 
The smoother is the interface, the weaker is the effect of the mesh.
Yet, a very wide domain wall requires larger 
systems to study the critical behaviour.
To choose a reasonable value of $\delta \approx \sqrt{c/\epsilon_0}$,
we perform two numerical experiments.

\begin{figure}[t!]
\includegraphics[width=0.95\columnwidth]{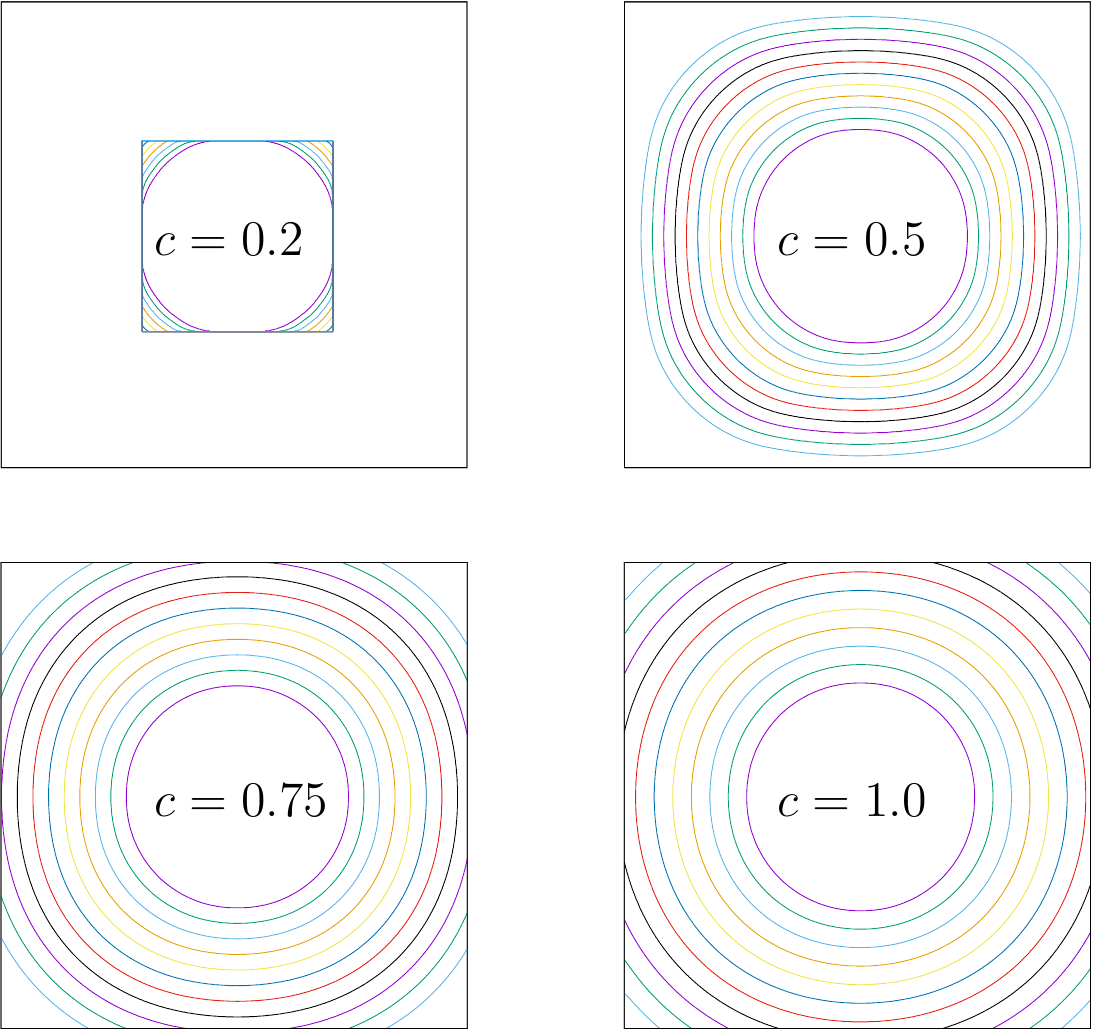}
\caption{Blowing an initially circular domain with a constant field $h=0.1$, for $\epsilon_0=1$ and different values of $c$ as indicated, corresponding to different values of the DW width $\delta \approx \sqrt{c/\epsilon_0}$. Lines of different colors corresponds to DW configurations at regularly distributed times.
}
\label{fig:isotropy}
\end{figure}

First, we test the relevance of the anisotropic effect induced by 
the mesh by blowing an initially circular domain in the absence of disorder.
We seed a circular domain at the center of the system and directly solve the dynamics of Eq.(\ref{eq:phi4}) at a constant field $h=0.1$,
for different DW widths by fixing $\epsilon_0=1$ and varying $c$.
In Fig.\ref{fig:isotropy} we observe that small values of $c$ 
($c=0.2$ and $c=0.5$) lead to an anisotropic growth that deviates 
from  the circular shape.
Further, one can see that the growth mean velocity is changing 
due to the mesh effect, since concentric lines correspond to 
regularly distributed times and they come closer to each other.
For $c=0.2$ the artificial periodic pinning is even able to pin the interface into a square metastable configuration.
Fortunately, at larger values of $c$ the undesired anisotropic 
effect rapidly vanishes ($c=0.75$ displays circular configurations and for
$c=1$ we cannot detect any sign of anisotropy). 
Note that in the latter case, $\delta \approx 1$, 
a width comparable with the lattice spacing. 
The uncorrelated disorder increases system isotropy  because 
rough DWs cannot coherently couple to the underlying 
periodic potential.

\begin{figure}[t!]
\includegraphics[width=1\columnwidth]{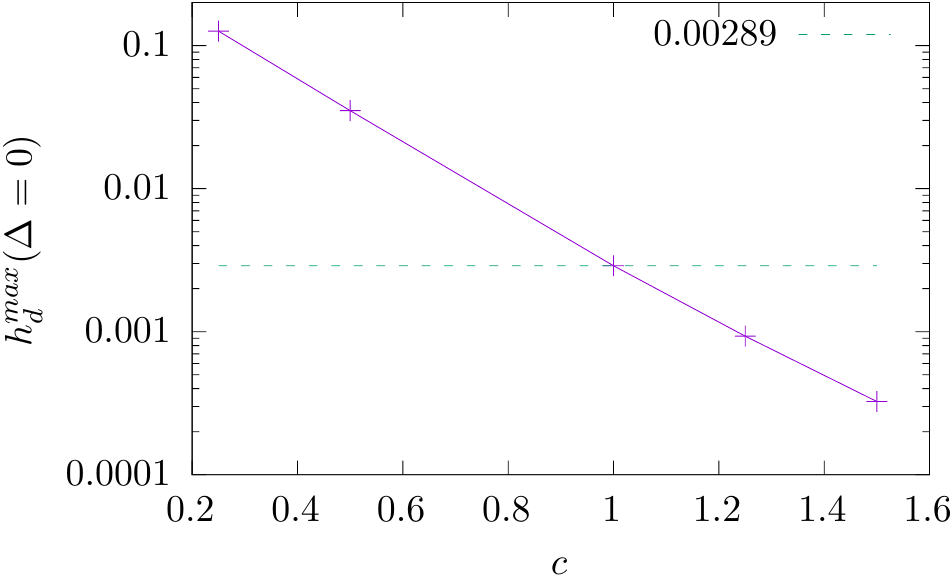}
\caption{
Maximum depinning field $\hd^{max}$ in the absence of disorder ($\Delta=0$) due to 
the effect of the square numerical mesh, vs the order parameter elastic constant $c$.
The dashed-line indicates $\hd^{max}$ for the parameters used to obtain our results 
in the main text.
}
\label{fig:hdgrid}
\end{figure}

To go further, we also compute the pinning field $\hd$ induced purely 
from the mesh in the absence of disorder ($\Delta=0$)
and compare it with the whole range of $\hd$ values that 
we analyzed in Fig.\ref{fig:hdvsDelta} for different disorder strengths $\Delta$.
As for the case $c=0.2$, $h=0.1$,
the maximum pinning field $\hd^{\max}(\Delta=0)$ is associated
to a square DW.
In Fig.\ref{fig:hdgrid} we show $\hd^{\max}(\Delta=0)$,
computed using our exact algorithm, as a function of the
elastic constant $c$.
We can observe that it decreases approximately exponentially 
with increasing $c$, (i.e increasing the DW  
width). 
In particular, for the choice $c=1$ that we adopted to obtain 
most of the results in our work, we get 
$\hd^{max}(\Delta=0)\approx 0.00289$, much smaller than 
the critical field obtained for the weakest disordered 
considered ($\Delta=0.1$). 
Therefore, we conclude  the mesh pinning does not affect our results. For even smaller $\Delta$ values, 
it may be necessary to increase the DW's intrinsic 
width $\delta \sim \sqrt{c/\epsilon_0}$, 
either increasing $c$ or decreasing $\epsilon_0$. 
Fig.\ref{fig:hdgrid} can be used as a general guide for that. 

In summary, we find that anisotropic effects can be kept under
control for any disorder strength. 
Our method thus constitutes an essential improvement over models 
of discrete scalar fields:
In the 2D-RFIM DWs are {\it always} sharp.
In particular, in the limit of weak disorder, square domains are not 
avoidable already at $\Delta \sim 1$~\cite{Ji1991, Koiller1992}. 
Fig.\ref{fig:hdvsDelta} shows that these kind of effects are absent 
in our simulations.
Therefore, our approach allows to test the prediction of the 
elastic theory in isotropic depinning.

\end{document}